\documentclass[10pt,aps,prl,twocolumn,superscriptaddress,longbibliography]{revtex4-2}
\usepackage{hyperref}
\usepackage{amsmath,amssymb,graphicx,bm}
\usepackage{lipsum}
\usepackage[utf8]{inputenc}
\usepackage[dvipsnames]{xcolor}
\usepackage{braket}

\immediate\write18{texcount -inc -sum main.tex > count.txt}

\def\iqo {\affiliation{Institut f\"ur Quantenoptik, Leibniz Universit\"at Hannover, Welfengarten 1, 30167 Hannover, Germany}}

\def\hub {\affiliation{Institut f\"ur Physik, Humboldt-Universit\"at zu Berlin, Newtonstra{\ss}e 15, 12489 Berlin, Germany}}

\begin{document}

\title{Universal Fold Caustics in Anharmonic Matter-wave Lenses}

\author{Joachim Guyomard}
\email[To whom correspondence should be addressed.\\]{guyomard@iqo.uni-hannover.de}
\iqo
\author{Julia Pahl} \hub
\author{Ernst Rasel} \iqo
\author{Naceur Gaaloul}\email{gaaloul@iqo.uni-hannover.de} \iqo

\date{\today}

\begin{abstract}

Caustics provide a universal connection between classical ray focusing and structured wave patterns. 
We theoretically investigate this connection in non-linear matter waves focused by a short-duration anharmonic lens acting on a Bose-Einstein condensate. 
A semi-classical description shows how the leading cubic aberration transforms the harmonic focal point into a fold caustic, whose local wave field is described by an Airy canonical integral. 
Three-dimensional Gross-Pitaevskii simulations show that mean-field interactions modify the near-focus density and longitudinal momentum distribution, producing a secondary momentum-space feature absent from the non-interacting evolution. 
We quantify the information about the effective interaction strength carried by the $1D$ distribution using classical Fisher information and maximum-likelihood estimation.
We obtain a relative uncertainty of $2.7\times10^{-3}$ for a sample of $10^5$ detected points along $z$. 
These results establish a framework for studying interaction-driven departures from single-particle caustic universality and promote matter-wave lenses as a sensitive probe of mean-field interactions.

\end{abstract}

\maketitle

\section{Introduction}

In optical systems anharmonicities and aberrations are described by the caustic formalism~\cite{Berry1980}.
In geometrical optics, caustics arise when neighbouring rays converge and the ray density becomes singular.
Wave optics regularizes those singularities into finite diffraction patterns whose local structure is governed by universal canonical forms~\cite{Berry1980}.
In particular, the canonical form near a generic fold caustic is described by an Airy function\cite{Airy1838}. 
Thus, a singularity predicted by ray optics gives rise to a characteristic and physically measurable interference pattern.

The same mechanism can be applied to matter waves.
Bose-Einstein Condensates (BEC) provide coherent matter-waves which spatial and momentum distributions can be shaped with techniques inherited from optical imaging and interferometry~\cite{Andrews1997,Mewes1997}.
Matter-wave caustics have been investigated in a range of settings, building on the broader theory of wave catastrophe~\cite{Chalker2009,Simula2013,Mossman2021,efremidis2013accelerating}.
For a condensed Bose gas, the question is therefore not whether an Airy pattern can occur, but under which conditions a generalized, anharmonic lens realizes the universal fold regime.
A further question is the extent the many-body interactions modify this pattern in a realistic system, and whether the resulting dynamics can be exploited for quantum sensing.

Atom-chip generated magnetic lenses provide a promising platform to address those questions.
Such lenses are commonly used for delta-kick or the controlled collimation of BECs~\cite{Arnold2002,Kovachy2015, Deppner2021, gaaloul2022,Szigeti2020,Corgier2021}.
Although the harmonic approximation captures their leading focusing dynamics, realistic magnetic lenses exhibit spatial anharmonicities even for state-of-the-art experiments~\cite{Deppner2021}.
We leverage these anharmonicities to propose a controlled route to investigate wave caustics.
In particular, we show, in a realistic system, how an anharmonic magnetic lens can transform the harmonic focal point into a fold caustic with an Airy diffraction pattern.

In this work, we derive a semi-classical description of an anharmonic matter-wave lens and identify the dimensionless parameters controlling the validity of the Airy approximation. 
The characteristic Airy length is set by the cubic lens coefficient, while the higher order contribution and condensate envelope provide controlled corrections. 
We then include mean-field interactions within a Gross-Pitaevskii Equation (GPE) simulation and show that the density enhancement associated to the caustic amplifies non-linear matter-wave dynamics, producing a systematic departure from the single-particle Airy universality.
Finally, we quantify the information about the interaction strength carried by the resulting spatial fringe distribution and show that the caustic provides a novel, sensitive probe of the mean-field interactions.

\section{System and lens model}
We consider a system close to the setup of the experiment of reference~\cite{Deppner2021} and adopt its parameters to conduct this study.
In an atom-chip setup, current passing through coil wires generate magnetic fields that act on the atoms through their magnetic moments. 
By appropriately ramping the wire and coil currents, the required potential can be used to operate a magnetic lens on the atomic cloud.

We consider a separable lens potential in which the anharmonicity is along the orthogonal-to-the-chip axis $z$.
This approximation captures the experimental situation of the lens realized in~\cite{Deppner2021}.
\begin{equation}
    V(x,y,z)=\dfrac{m}{2}\left(\omega_x^2 x^2 +\omega_y^2y^2 + \omega_z^2z^2\left(1+\dfrac{z}{L_3}+\dfrac{z^2}{L_4^2}\right)\right)
\label{eq:lenspotential}
\end{equation} Coupling between transverse and longitudinal coordinates is neglected here. 
Even if this separable potential is used here, the analytical framework could be extended to non-separable potentials by expanding the relevant generating function around the caustic of interest.
The analytical framework does not rely on the specific values of $L_3$ and $L_4$, but solely assumes the ones of~\cite{Deppner2021} for numerical simulations.
\begin{figure}[b!]
    \centering
    \includegraphics[width=0.9\linewidth]{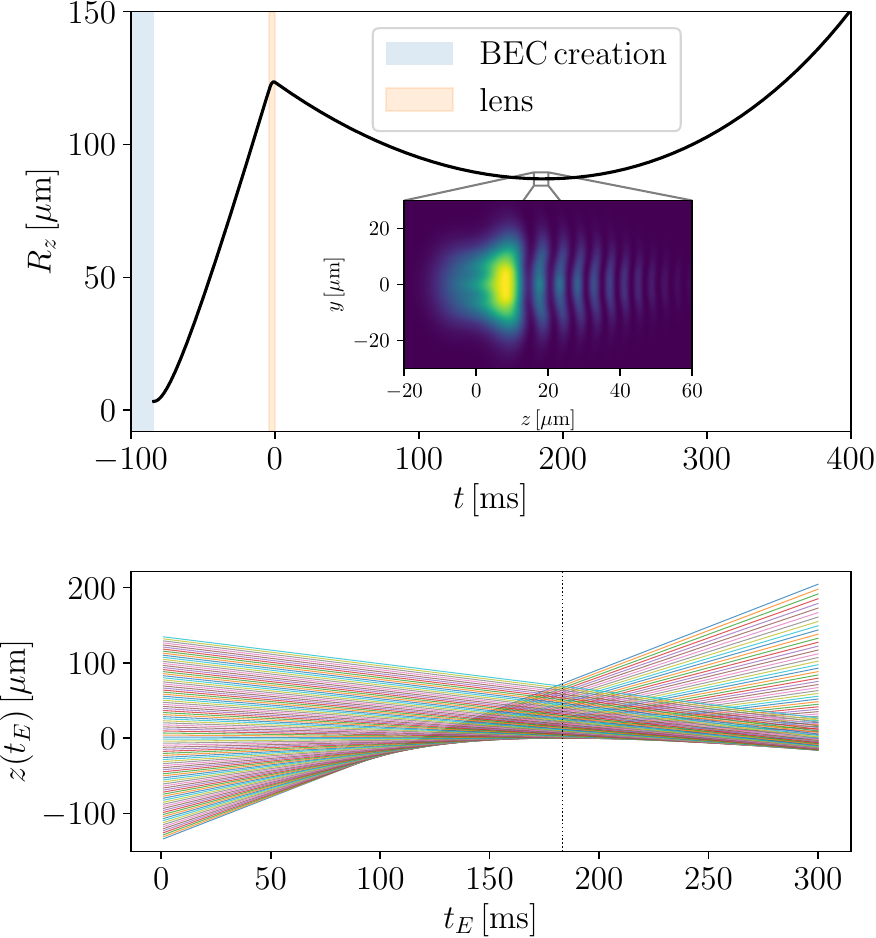}
    \caption{The top plot shows the evolution of the BEC size (TF-radius) during the studied sequence. Our study starts at the beginning of the orange span (lens). The inset shows the transverse probability density of the simulated BEC cloud given the initial parameters extracted from~\cite{Deppner2021}.
    The bottom plot shows the evolution of the classical ray family forming the inner-classical trajectories inside the BEC wave-function after being lensed.}
    \label{fig:Sequence_and_2D_plot}
\end{figure}

As initial state we will take a BEC in the Thomas-Fermi (TF) regime, with initial parameters extracted from~\cite{Deppner2021}, giving $N=10^5 $, TF radii $R_i=\{93.1;131.39;133.26\}\,\mathrm{\mu m}$ and expansion velocities $v_i = \{0.809;1.416;1.597\}\,\mathrm{mm/s}$.
We consider the dynamics from the instant at which the lens is applied, and consider it acting for a duration $t_L$, after which the lens is switched off and the cloud evolves freely for a time $t_E$.

The top plot of Fig.\ref{fig:Sequence_and_2D_plot} depicts a typical size evolution of a BEC experiencing a magnetic lens, with parameters taken from~\cite{Deppner2021}. 
As collimation has been reported to happen for a lens duration of $t_L=2.42\,\mathrm{ms}$ in~\cite{Deppner2021}, we have applied here a lens duration of $t_L=3.9\,\mathrm{ms}$ to ensure focusing.
The blue span represents the experimental work to produce the BEC, followed by free expansion during which the interaction energy is converted to kinetic energy setting the BEC in TF regime. 
The orange span marks the lensing stage and thus the beginning of our study. The following free expansion exhibits as expected a focusing dynamic. 
The inset shows a density plot the deformed BEC simulated for the lens parameters extracted from~\cite{Deppner2021}, exhibiting how anharmonicity during the lensing can produce a non-trivial shape in the BEC during focusing.

For the following theoretical investigations, we first consider the dilute case in which the BEC dynamic is described by the single-particle wave-function, allowing us to perform a completely analytical $1D$ semi-classical model predicting the emergence of interference fringes near the fold caustic.
We then numerically solve the $3D$ GPE using a symmetric Strang split-step method~\cite{strang_split_step,Seckmeyer2026} to investigate how the mean-field interactions modify the caustic structure. 


\section{Single-particle analytical treatment}

The anharmonicity of the lens requires an extension of the standard ray-transfer description of delta-kick potentials. We first consider the harmonic limit, for which the propagation through a lens of duration $t_L$ is described by the transfer matrix~\cite{Gerrard1994}
\begin{equation}
    \begin{pmatrix}
        \mathfrak{z}_{t_L}\\
        \mathfrak{v}_{t_L}
    \end{pmatrix} = 
    \begin{pmatrix}
     \cos{(\omega_zt_L)}& \sin{(\omega_z t_L)}/\omega_z \\
     -\omega_z \sin(\omega_z t_L)& \cos{(\omega_z t_L)}
    \end{pmatrix}
    \begin{pmatrix}
        z_i \\
        v_i
    \end{pmatrix}   
\label{eq:paraxial_ABCD}
\end{equation}
where $\omega_z = 2\pi f_z$, with $f_z$ the lens frequency of the trap along $z$.
We use here gothic letters for describing the propagation after the lens.

In contrast to the usual optical treatment, the expanding condensate enters the lens with a correlated position-momentum distribution. During free expansion, a BEC in the TF regime develops a self-similar velocity field~\cite{Castin1996,Dalfovo1999}
\begin{equation}
    v(z,t)=\frac{\dot b_z(t)}{b_z(t)}z_i
=\frac{\dot R_z(t)}{R_z(t)}z_i 
\end{equation}
We therefore characterize the initial phase-space correlation at the entrance of the lens with a chirp
\begin{equation}
    \alpha:=v_{z,0}/R_{z,0},
\qquad
v_i=\alpha z_i
\end{equation}
where $R_{z,0}$ and $v_{z,0}$ denote the TF radius and the corresponding expansion velocity along the $z$-axis immediately before lensing. Giving
\begin{equation}
    \begin{cases}
        \mathfrak{A}:=\cos{(\omega_zt_L)} +\alpha \sin{(\omega_z t_L)}/\omega_z \\
        \mathfrak{B}:=-\omega_z \sin(\omega_z t_L) + \alpha \cos{(\omega_z t_L)}
    \end{cases} \Rightarrow 
    \begin{cases}
        \mathfrak{z}_{t_L}=z_i\,\mathfrak{A} \\
        \mathfrak{v}_{t_L}=z_i\, \mathfrak{B}
    \end{cases}
\label{eq:lensed_ABCD}
\end{equation}
\begin{equation}
    \mathfrak z_h(t_E) = \mathfrak{z}_{t_L} + \mathfrak{v}_{t_L} t_E = z_i\,(\mathfrak{A}+\mathfrak{B}\,t_E) 
    \label{eq:paraxial_z}
\end{equation}
This set of equations gives directly the expected focal time $t_f$ for the family of rays as $t_f = -\mathfrak{A}/\mathfrak{B}$~\cite{Gerrard1994}.
Thus, for known $\omega_z$ and $t_L$, the measured focal time directly determines the initial expansion rate $\alpha$. We can now extend this simple model to our anharmonic lens. Equation~\eqref{eq:paraxial_z} describes the harmonic part of the lens, and the anharmonic contribution will be considered as a perturbation $\delta$. From the expression of our potential of Eq.~\eqref{eq:lenspotential}, we can note the two coefficients
\begin{equation}
    \beta :=\dfrac{1}{2} \dfrac{\omega_z^2}{L_3} 
    \qquad
    \text{and} 
    \qquad
    \gamma := \dfrac{1}{2}\dfrac{\omega_z^2}{L_4^2},
\end{equation}
thus one can deduce the acceleration implied by our potential as 
\begin{equation}
    \ddot{z} = - \omega_z^2 z - 3 \beta z^2 - 4 \gamma z^3
\label{eq:anharmonic_acceleration}
\end{equation}
To first order in $\beta$ and $\gamma$, the position correction at an intermediate time $\tau$ given by Green function formalism of harmonic oscillator is
\begin{equation}
\begin{aligned}
\delta z(\tau)
=&\,-\dfrac{3\beta}{\omega_z}
\int_0^{\tau}
\sin(\omega_z(\tau-t))\,\mathfrak z_h(t;z_i)^2\,\mathrm{d}t\\
&\,-\dfrac{4\gamma}{\omega_z}
\int_0^{\tau}
\sin(\omega_z(\tau-t))\,\mathfrak z_h(t;z_i)^3\,\mathrm{d}t
\end{aligned}
\label{eq:perturbation_z}
\end{equation}
Thus during the lensing the perturbation will add an extra velocity contribution in the phase-space
\begin{equation}
\begin{aligned}
\delta v(\tau)
=&\,-3\beta\int_0^{\tau}
\cos(\omega_z(\tau-t))\,\mathfrak z_h(t;z_i)^2\,\mathrm{d}t\\
&\,-4\gamma\int_0^{\tau}
\cos(\omega_z(\tau-t))\,\mathfrak z_h(t;z_i)^3\,\mathrm{d}t
\end{aligned}
\label{eq:perturbation_v}
\end{equation}
Those integrals are closed and admit analytic solutions which should be retained in the general case, however in the short-lens approximation $(\omega_zt_L\ll1\, \&\, \vert\alpha\vert t_L\ll1)$, one finds
\begin{equation}
\begin{aligned}
\delta v(t_L)
&\sim -3\beta t_L z_i^2 -4\gamma t_L z_i^3 + O(t_L^2)\\
\delta z(t_L)
&\sim -\frac{3}{2}\beta z_i^2t_L^2 -2\gamma z_i^3t_L^2 + O(t_L^3)
\end{aligned}
\label{eq}
\end{equation}
We will write $k_3$ the coefficient of $z_i^2$ and $k_4$ the coefficient of $z_i^3$, and obtain the perturbed velocity at the lens exit
\begin{equation}
    \mathfrak{v}_{t_L} \simeq z_i \mathfrak{\,B} + k_3z_i^2 + k_4 z_i^3\;;\;k_3=-3\beta t_L\,,\,k_4=-4\gamma t_L
    \label{eq:perturbed_v}
\end{equation}
We retain the leading anharmonic correction to the exit velocity $O(t_L)$ and neglect the corresponding exit-position correction $O(t_L^2)$. 
This approximation is used only when the latter produces a negligible change in the propagated ray map over the times and initial-coordinate range considered.
This leads to the classical ray-family 
\begin{equation}
    z(t_E) \simeq z_i\,(\mathfrak{A} + \mathfrak{B}\,t_E) + t_E(k_3z_i^2 + k_4z_i^3)
\label{eq:z_extended_ABCD}
\end{equation}
and thus given the caustic coalescence criteria $\partial_{z_i} z(t_E)=0$, yields
\begin{equation}
    \Rightarrow t_{c}(z_i) =-\dfrac{\mathfrak{A}}{\mathfrak{B} +2k_3z_i + 3k_4z_i^2} 
\label{eq:focal_caustic}
\end{equation}
In the harmonic limit, all rays share the same focal time. Anharmonicity instead renders the focal time position-dependent, transforming the focal point into a caustic envelope~\cite{Berry1980}. Thus making the harmonic focal time the natural expansion point at which the quadratic phase vanishes, while the anharmonic system possesses a spatially extended fold caustic.

We have represented the evolution of this classical ray-family on the bottom plot of Fig.\ref{fig:Sequence_and_2D_plot}. One can see the folding of ray-family on itself creating the standard cubic caustic. 
The vertical line shows the harmonic focal time, which corresponds to the instant where the instant where the dispersion of the ray-family is minimal.

We then seek the corresponding semi-classical phase. From the ray map we construct a local generating function $S_L(z_i)$ whose spatial derivative gives the exit momentum after the lens. In the semi-classical approximation~\cite{Littlejohn1992},
\begin{equation}
\psi_L(z_i)=E(z_i)\exp\left(\frac{i}{\hbar}S_L(z_i)\right)
\end{equation}
with the generating function :
\begin{equation}
\frac{\partial S_L}{\partial z_i}=m\mathfrak v_{t_L}
\frac{\partial\mathfrak z_{t_L}}{\partial z_i}
\end{equation}
And $E(z_i)$ the envelope of the wave-function.
Using Eq.\eqref{eq:z_extended_ABCD}, this gives, up to an additive constant,
\begin{equation}
  S_L =  m\mathfrak A\left(
\frac{\mathfrak B}{2}z_i^2+
\frac{k_3}{3}z_i^3+
\frac{k_4}{4}z_i^4
\right)
\end{equation}
We can now propagate the lensed wave-function thanks to the standard free evolution propagator and find 
\begin{equation}
\begin{aligned}
    \psi_{t_E}(z) =& \sqrt{\dfrac{m}{i2\pi\hbar t_E}\left\vert\dfrac{\partial\mathfrak{z}_{t_L}}{\partial z_i}\right\vert}\int \mathrm{d}z_i\,E(z_i)\\
    &\exp{\left(\dfrac{i}{\hbar}\left(S_L(z_i) + \dfrac{m}{2t_E}(z-\mathfrak{z}_{t_L})^2\right)\right)}
\end{aligned}
\label{eq:semiclassical_free_propagation}
\end{equation}
This expression yields the semiclassical non-interacting wavefunction after the anharmonic lens. 
It shows that a single TF radius is insufficient to describe the spatially varying phase generated by the anharmonic lens. 
Since the interference fringes are encoded in the phase of the wavefunction, we restrict our attention to computing
\begin{equation}
\begin{aligned}
    \Phi(z_i)=&\dfrac{m}{2t_E}z^2 - \dfrac{m\mathfrak{A}}{t_E}zz_i \\ &+ m \mathfrak{A}\left(\dfrac{1}{2}\left(\mathfrak{B}+\dfrac{\mathfrak{A}}{t_E}\right)z_i^2 + \dfrac{k_3}{3}z_i^3 + \dfrac{k_4}{4}z_i^4\right) 
\end{aligned}
\label{eq:exact_SC_phase}
\end{equation}
Evaluating the phase near the harmonic focus $t_f$ the quadratic term disappears. Assuming the caustic is centred at $z_i=0$, this yields
\begin{equation}
    \psi_{t_f}(z)=C\int \mathrm{d}z_i\,E(z_i)\exp{\left(i\dfrac{m\mathfrak{A}}{\hbar}\left(-\dfrac{z}{t_E}z_i + \dfrac{k_3}{3}z_i^3+\dfrac{k_4}{4}z_i^4\right)\right)} 
\label{eq:semi_analytic}
\end{equation}
where $C$ contains the propagation prefactor, the constant Jacobian factor, and the global phase. At a fold caustic, the leading uniform approximation is described by an Airy integral. A quartic term can become important in a higher-order degeneracy, such as a cusp, where the appropriate canonical integral is the Pearcey integral~\cite{Airy1838,Pearcey1946,Berry1980}.
We truncate the phase at cubic order, neglecting the quartic contribution, such that 
\begin{equation}
    \Phi(z_i)\simeq m\mathfrak{A}\left(-\dfrac{z}{t_E}z_i + \dfrac{k_3}{3}z_i^3\right) 
\end{equation} 
We can now define the canonical scale for the Airy function describing our problem~\cite{Wright1988, Vallee2004}:
\begin{equation}
    \ell = \left(\dfrac{\hbar}{m\vert\mathfrak{A} k_3 \vert}\right)^{1/3} \Rightarrow \ell \propto t_L^{-1/3}\;;\; u:=\dfrac{z_i}{\ell}
\label{eq:characteristic_scale}
\end{equation}
Given this characteristic length, we can estimate the ratio between the cubic and quartic contributions as
\begin{equation}
    \epsilon_4 = \left\vert\dfrac{3k_4}{4k_3}\ell \right\vert \simeq \left\vert\dfrac{L_3}{L_4^2}\ell \right\vert\ll1,
\end{equation}
For the parameters given in~\cite{Deppner2021} we find $\epsilon_4\simeq4.6\times 10^{-3}$ which justifies the cubic truncation for this experimental consideration. Furthermore, we can also propose a criterion for a slowly varying envelope as 
\begin{equation}
    \epsilon_\mathrm{env}=\dfrac{\ell^2}{R_z^2}\ll1
\end{equation}
For the parameters considered~\cite{Deppner2021}, we find $\epsilon_\mathrm{env}\simeq5.9\times 10^{-2}$,
we can therefore approximate $E(z_i)$ by its local value $E(0)$, which reduces our integral to
\begin{equation}
\begin{aligned}
    \psi_{t_f}(z)\propto \int \mathrm{d}u\,\exp{\left(i\left(\dfrac{u^3}{3}+\zeta u\right)\right)}\\
    \mathrm{with :}\:\:\zeta:=-\mathrm{sign}(\mathfrak{A}k_3)\dfrac{z}{\kappa_\ell}\; ;\; \kappa_\ell = \dfrac{\hbar t_f}{m\mathfrak{A}\ell}
\end{aligned}
\label{eq:airy_integral}
\end{equation}
This is the canonical integral representation of the Airy function in Fourier space~\cite{Vallee2004}, which gives
\begin{equation} 
    \psi_{t_f}(z) \propto \mathrm{Ai}(\zeta) \Rightarrow n_{t_f}(z) \propto \vert\mathrm{Ai}(\zeta)\vert^2 
\label{eq:analytical_airy}
\end{equation}
An even better approach than solely give the Airy fold is to compute the precise wave-function, since we are looking for the $1D$ projection of the $3D$ BEC density.
For the analytical calculation, we construct an effective one-dimensional initial amplitude whose squared modulus reproduces the axial marginal of the three-dimensional TF density. We propagate this amplitude using the longitudinal single-particle Hamiltonian. This construction is an effective model of the $1D$ marginal.
\begin{equation}
\begin{aligned}
    \psi_{t_f}(z)=&\sqrt{\dfrac{m}{i2\pi\hbar t_f}\dfrac{15}{16R_z} \left\vert\dfrac{\partial\mathfrak{z}_{t_L}}{\partial z_i}\right\vert}\int_{-R_z}^{R_z}\left(1-\dfrac{z_i^2}{R_z^2}\right)\\
    &\exp\left(i\dfrac{m\mathfrak{A}}{\hbar}\left(\dfrac{k_3}{3}z_i^3 -\dfrac{z}{t_f}z_i\right)\right)\mathrm{d}z_i
    \\
    =&\sqrt{\dfrac{m}{i2\pi\hbar t_f}\dfrac{15}{16R_z}\left\vert\dfrac{\partial\mathfrak{z}_{t_L}}{\partial z_i}\right\vert}\\ 
    &\int\ell\mathrm{d}u\left(1-\dfrac{\ell^2 u^2}{R_z^2}\right)\exp{\left(i\left(\dfrac{u^3}{3} + \zeta u\right)\right)}
\end{aligned}
\end{equation}
Here we have extended the bound of the integral from the TF radius to infinity, which allows us to use the Airy function properties~\cite{Vallee2004}
\begin{equation}
\begin{aligned}
    \int du \exp(i(u^3/3 + \zeta u))=&\,2\pi\mathrm{Ai}(\zeta)\\
    \zeta\mathrm{Ai}(\zeta) =&\, \mathrm{Ai}''(\zeta)\\
\partial_{\zeta}^2\exp(i(u^3/3 + \zeta u))=&\,-u^2\exp(i(u^3/3 + \zeta u))\\
\Rightarrow\int du \:u^2\exp(i(u^3/3 + \zeta u))=&\,-2\pi\mathrm{Ai}''(\zeta)=-2\pi\zeta \mathrm{Ai}(\zeta)
\end{aligned}
\end{equation}
We can finally find an explicit analytic expression for the wave-function at the harmonic focal time.
\begin{equation}
    \psi_{t_f}\sim \sqrt{\dfrac{m}{i2\pi\hbar t_E}\dfrac{15}{16R_z} \mathfrak{A}}\,2\pi\ell\mathrm{Ai}(\zeta)\left(1+\ell^2 \dfrac{\zeta}{R_z^2}\right)
\label{eq:anl_wave_function}
\end{equation}
This analytical form is only valid inside the cloud $[-R_z;R_z]$, as the Airy function has an infinite oscillatory tail which physical normalization is only restored by the cloud finite size.

Varying $t_L$ changes the physical scale of the caustic through the transfer coefficients and the cubic phase coefficient. After rescaling the output coordinate by the corresponding Airy length, the leading-order fold profile is independent of these parameters. Residual differences arise from the finite initial envelope, the quartic phase, and deviations from the local fold approximation.
This means that the fold caustic is the universal result for the single-particle wave-function obtained by cubic lensing.
The structure appears in the focus independent of the duration of the delta kick given the approximation of thin lens valid.

Finally this model can give the expression of the generated $1D$ wave function in our anharmonic lens for any $t_E$ given the integral Eq.\eqref{eq:semiclassical_free_propagation}
to be performed numerically, with $\mathfrak{v}_{t_L}$ and $\mathfrak{z}_{t_L}$ containing the general correction given by Eq.(\ref{eq:perturbation_z};\ref{eq:perturbation_v}) and the correct envelope.
This framework can also be extended for a BEC with an initial position $z_0$ inside the lens.

\section{Results}
\subsection{Single-particle Wave-function Collapse and Interaction-induced dynamic}

\begin{figure}[t!]
\centering
    \includegraphics[width=0.9\linewidth]{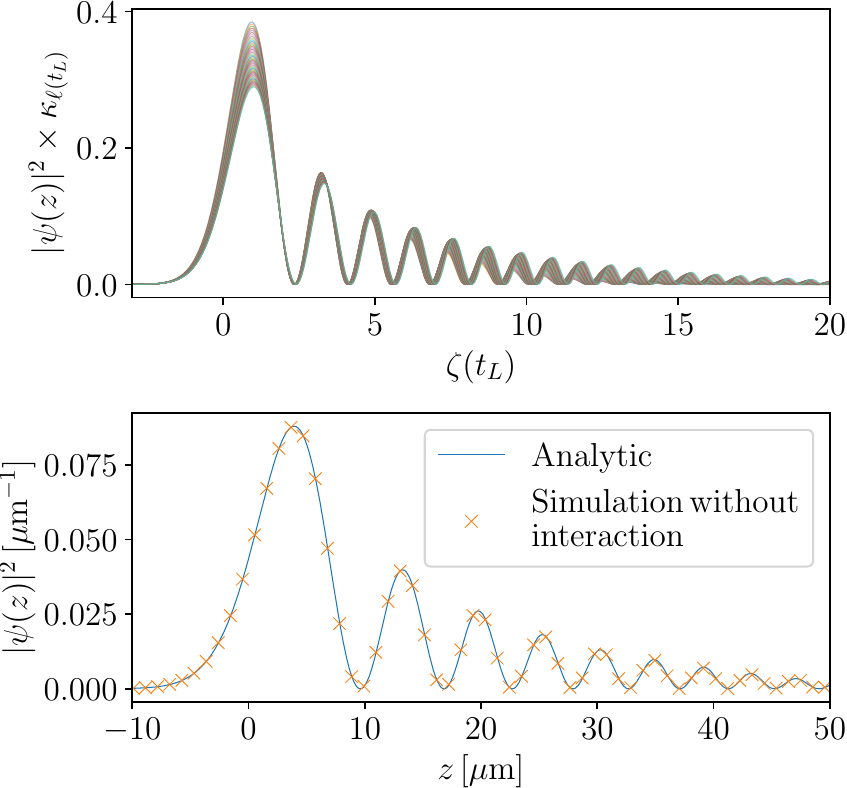}
    \caption{
    The top plot shows single-particle wave-functions with initial parameters for the BEC and lens extracted from~\cite{Deppner2021} for $50$ values of $t_L$ between $3$ and $8$ $\mathrm{ms}$. Each wave-function was rescaled by its corresponding $\kappa_\ell$.
    On the bottom plot the orange crosses are the $1D$ marginal of the wave-function given by our $3D$ simulation. The blue curve is our analytical prediction of equation~\eqref{eq:anl_wave_function}. We have chosen $t_L=3.9\,\mathrm{ms}$, and the parameters are taken from~\cite{Deppner2021}.}
    \label{fig:collapse_comparison_simu_anl}
\end{figure}

We first exhibit the emergence of the fold caustic in the non-interacting limit.
The top plot of Fig.\ref{fig:collapse_comparison_simu_anl} shows the wave-functions obtained for fifty different lens durations, $t_L\in[3;8]\,\mathrm{ms}$.
Each wave-function having been rescaled with its corresponding scale $\kappa_\ell$, they collapse onto a common profile.
This collapse shows the universal form of the expected local fold diffraction pattern for our matter-wave.

The bottom plot of Fig.\ref{fig:collapse_comparison_simu_anl} compares the analytical prediction given by \eqref{eq:anl_wave_function} and the $1D$ marginal density along $z$ obtained from the non-interacting Schrödinger simulation for $t_L=3.9\,\mathrm{ms}$.
The excellent agreement supports the validity of the Airy approximation for the parameters considered.
Residual differences arise from corrections given by both the neglected quartic contribution and the finite condensate envelope.

\begin{figure}[b!]
\centering
    \includegraphics[width=0.9\linewidth]{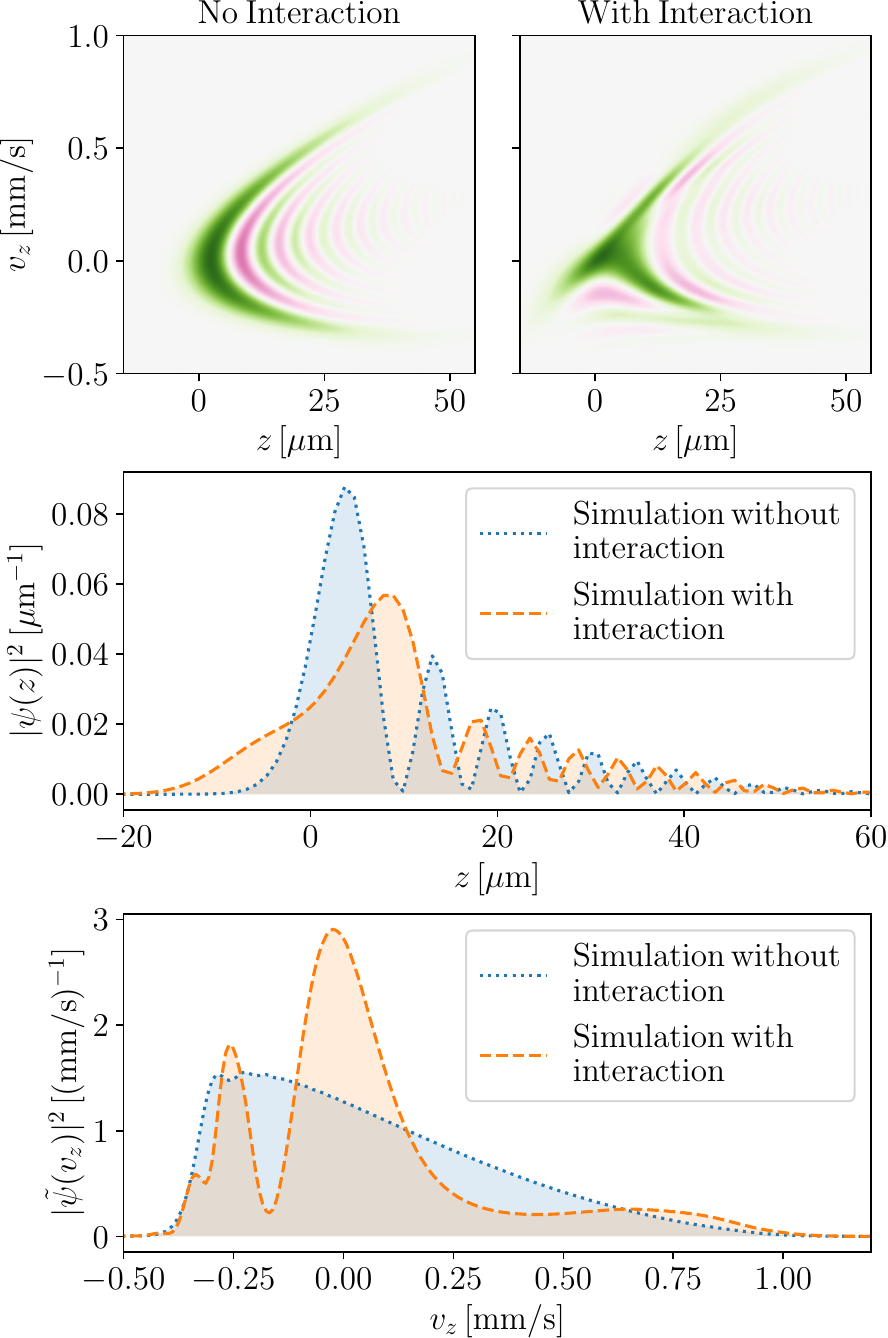}
    \caption{The top plot shows the Wigner distributions of the BEC reduced to the $z$-axis, and the function is normalized such that the distribution goes between $-1$ (pink) and $1$ (green). The left plot is made without interaction, while the right one is made for standard mean-field of $^{87}\mathrm{Rb}$.
    The two plots below depict the marginal integration of the Wigner distribution showing the wave-function in both position and velocity space. The blue curves are from the non-interacting simulation, while the orange one are for standard $^{87}\mathrm{Rb}$ mean-field interaction.}
    \label{fig:wigner_and interaction}
\end{figure}

The caustic can also be understood in terms of interference between inner classical trajectories of the wave-function. 
Indeed as illustrated by the ray map of Fig.\ref{fig:Sequence_and_2D_plot} and the non-interacting phase-space distribution of Fig.\ref{fig:wigner_and interaction}, two inner classical trajectories with different momenta can reach the same position at a given observation time.
Their coherent superposition produces the oscillatory fringes characteristic of the fold.
The spatial scale $1/\kappa_\ell$ characterizes the Airy pattern and therefore the separation between the fringes.

By neglecting interactions, we isolate the single-particle anharmonicity of the lens from the many-body nonlinearity associated with the condensate mean-field. 
This deviation captured by the GPE simulation shown on Fig.\ref{fig:wigner_and interaction} provides a direct signature of mean-field dynamics. 
This figure represents the Wigner distribution and its marginals for the simulated wave-function with and without interaction. The simulation being $3D$ we have here reduced it to $1D$ using the density matrix defined as~\cite{cohen-tannoudji1973a}
\begin{equation}
    \rho_z(z,z')=\int \mathrm{d}x\,\mathrm{d}y\,\psi(x,y,z)\psi^*(x,y,z')
\label{eq:density_matrix}
\end{equation}
Defining then our Wigner function~\cite{wigner1932} as 
\begin{equation}
    W(z,p_z)=\dfrac{1}{2\pi\hbar}\int \mathrm{d}s\,\rho_z(z+s/2,z-s/2)e^{-ip_zs/\hbar}
\end{equation}

The resulting Airy profile is intrinsically asymmetric, with an oscillatory tail on one side and rapid decay on the other, thus the atomic density is unevenly distributed along the $z$-axis.
The strong density gradients generated near the caustic amplify the role of mean-field interactions. 
In our simulations, we observe a distinct velocity component emerging from the main distribution near the harmonic focal time, indicating the ejection of a fraction of the condensate from the high-density region. 

This phase-space signature provides a direct visualization of the interaction-induced deviation from the non-interacting semi-classical dynamics.
A weakly interacting BEC behaving mostly linearly away from focus becomes non-linear precisely because the caustic concentrates the matter-wave.
While the interaction-less distribution is well described by the single-particle ray picture, the strong density enhancement near the caustic increases the local mean-field energy and thus amplifies non-linear dynamics. 

To estimate the importance of the accumulated mean-field phase, we introduce a dimensionless quantity analogous to the non-linear $B$ integral used in non-linear optics~\cite{hache2017optique}
\begin{equation}
    B_\mathrm{MF} = \dfrac{g}{\hbar}\int_\mathrm{relevant} n_\mathrm{eff}(t) \,\mathrm{d}t\:\:\mathrm{with}\:\:n_\mathrm{eff}(t)\sim \dfrac{N}{\Omega(t)}
    \label{eq:B_integral}
\end{equation}
With $\Omega = \sqrt{\det \Sigma}$ where $\Sigma$ is the spatial covariant matrix of the distribution computed via the GPE simulation, this gives us a simple effective density.

As an illustrative comparison, we approximate $n_\mathrm{eff}$ constant over a $1\,\mathrm{ms}$ interval, and evaluate this $B_\mathrm{MF}$ integral during the lens, and during harmonic focus. 
This gives $B_\mathrm{MF}(\mathrm{lens})\simeq 0.05$ and  $B_\mathrm{MF}(\mathrm{focus})\simeq0.8$.
Even if these are approximated, this simple analogy illustrates the amplification of the mean-field interaction induced by the focusing.
Indeed the criterion $B\gtrsim 1$ often used to indicate a non-linear phase accumulation in optics, provides a useful qualitative analogy here.

We also evaluate the dilute-gas parameter given as $\chi =\sqrt{n_\mathrm{eff}(t) a_s^3} \ll 1$ where $a_s$ is the $s$-wave scattering length~\cite{LHY_correction}.
Using our previously introduced effective density near the harmonic focal we find $\chi\simeq1.6\times 10^{-3}$.
This parameter characterizing the scale of corrections beyond leading-order mean-field theory, justifying the mean-field approximation despite the strong non-linear evolution.

\begin{figure}[b!]
        \includegraphics[width=0.9\linewidth]{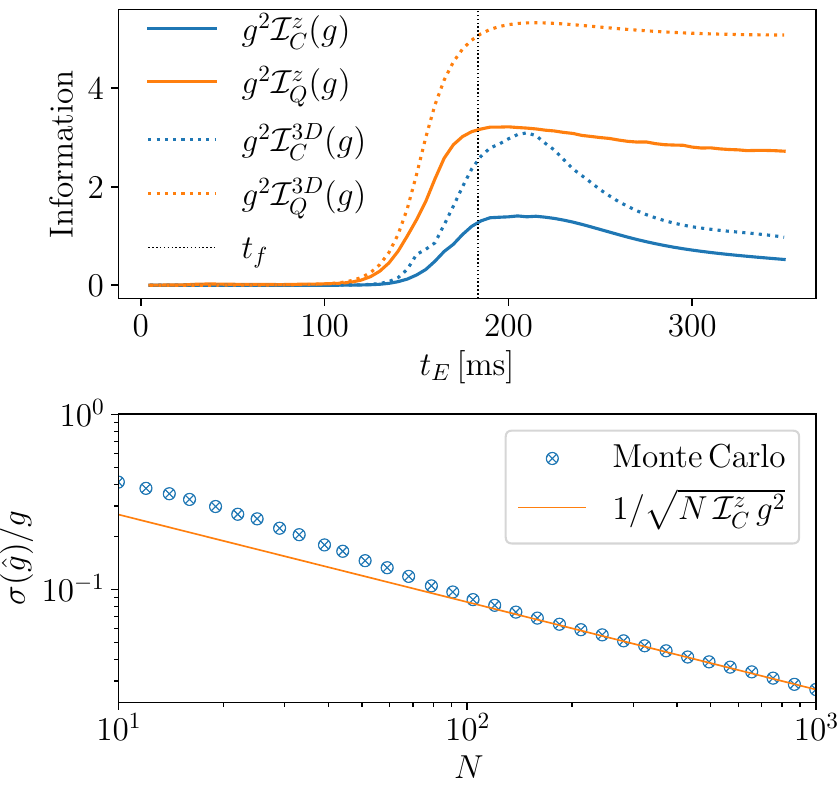}
        \caption{The top plot shows the Fisher information on the mean-field along $t_E$ for both the $3D$ distribution, dashed lines, and the $1D$ marginal along the $z$-axis, solid lines. The bottom plot presents Monte Carlo simulation for mean-field estimation over the $1D$ marginal along the $z$-axis.}
        \label{fig:fisher_info_and_MC}
    \end{figure}

\subsection{Mean-field sensing}

The interaction-induced modification of the position and momentum distributions suggests that the anharmonic lensing of matter waves may provide a sensitive probe of the mean-field dynamic inside the quantum gas.
We quantify this sensitivity by evaluating the classical Fisher information (CFI) carried by position measurement and comparing it with the quantum Fisher information (QFI).

Assuming a completely coherent pure state we compute the CFI after renormalizing the wave-function from $N$ to $1$ following reference~\cite{Fisher_info}
\begin{equation}
    \mathcal{I}_C(g) =\int \left(\partial_g \vert \psi_g(\Vec{r})\vert^2\right)^2 / \vert \psi_g(\Vec{r})\vert^2 \mathrm{d}\Vec{r}
\end{equation}
For the $3D$ QFI, under the same assumptions we compute 
\begin{equation}
    \mathcal{I}_Q(g) = 4 \left(\int  \vert\partial_g \psi_g(r)\vert^2 \mathrm{d}r - \left\vert\int\psi_g^*(r) \,\partial_g \psi_g(r)\right\vert^2\right)
\end{equation}
For the $1D$  QFI we use again our reduced density matrix Eq.\eqref{eq:density_matrix} and use the symmetric logarithmic derivative formalism, expressed in the eigenbasis of $\rho_z$~\cite{paris2009quantum}.
We use the standard definition $g=4\pi\hbar^2 a_s/m$ as the mean-field interaction coefficient.

The top plot of Fig.\ref{fig:fisher_info_and_MC} shows the evolution of the Fisher information on the mean-field coefficient carried over the focusing evolution for both the $3D$ distribution and the $1D$ marginal on the $z$-axis.
For the $3D$ distribution the Fisher information are not reaching its maxima at the harmonic focal time but a bit later.
This is consistent with the delayed emergence of the secondary momentum-space feature shown on Fig.\ref{fig:wigner_and interaction}. 
The interaction-induced modification of the distribution continues to evolve beyond the focus. 
The $1D$ marginal retains information about the coupling, yet as expected its content differs from the full $3D$ distribution.

We propose on the bottom plot of Fig.\ref{fig:fisher_info_and_MC} the evolution of the relative precision over the mean-field coefficient measured through coupled Monte-Carlo draws and maximum Likelihood estimator on the $1D$ projection of the BEC along the $z$-axis at $t_E = 210\,\mathrm{ms}$.
This makes it possible to experimentally measure along a single axis.
For each given $N$ the $\sigma(\hat{g})$ is the standard deviation given by $10.000$ repetitions of the random $N$ draws procedure, with our estimator $\hat{g}$ defined as
\begin{equation}
    \hat g=\underset{g_t\in\mathcal G_t}{\arg\max}[\log\mathcal L(g_t\mid z_1,\ldots,z_N)]
\end{equation}
With $\mathcal L$ the likelihood of a test value $g_t$ in our set $\mathcal G_t$ of $1000$ candidate values centred around $g$, given the $N$ position randomly drawn.
We show that the Cramer-Rao bound~\cite{Cramer_statistic} is quickly saturated as simply $100$ points randomly drawn in the probability distribution is already reaching the $1/\sqrt{N}$ regime.
Furthermore from this we can deduce that for $10.000$ points randomly drawn along the $1D$ $z$-axis distribution, we reach a $8.5\times10^{-3}$ relative precision on the mean-field dynamic, and for $100.000$ points we reach $2.7\times 10^{-3}$ relative precision.
As an illustration, assuming a perfect detection system for a cloud of $100.000$ atoms with a mean-field expansion around $1\,\mathrm{nK}$, for $10^4$ and $10^5$ sampled points one could have a precision around respectively  $8.5\,\mathrm{pK}$ and $2.7\,\mathrm{pK}$.

\section{Discussion}

In this article we have shown that a realistic anharmonic matter-wave lens provides a tunable realization of a universal fold caustic.
The resulting Airy structure acts as a non-linear amplifier of weak mean-field interactions, from which we can thus measure the mean-field coefficient inside a BEC cloud.
This study is highly realistic since the parameters are those of an existing experiment~\cite{Deppner2021}. It motivates further studies of an optimized configuration where cubic lensing could lead to competitive sensitivity in mean-field measurements in situations where expansion slopes will be limited by imaging artefacts such as collimated BECs in the pK regime or below.
The present simulations establish a route to estimating the effective mean-field interaction strength within the GPE model. An extension to beyond-mean-field parameters would require a corresponding dynamical model and a separate assessment of identifiability and sensitivity.
The information-theory analysis quantifies the sensitivity of axial position measurements to the interaction parameter under an idealized model. Establishing the experimental performance will require accounting for finite resolution, atom-number uncertainty, and other spurious parameters that go beyond this study.
Finally and given the parameters taken in this work, we demonstrate that already existing atom-chip experiments are able to create caustic behaviour. 
The combination of both the $\ell$ and $\kappa_\ell$ coefficients should be enough to give the fringe spacing of such caustics and thus check if the detection resolution of an experiment is sufficient to resolve the fringes. 
We are confident that this technique can also be implement in optical potential scheme, and painted potentials.

\section{Acknowledgements}
This work was funded by the Deutsche Forschungsgemeinschaft (German Research Foundation) under Germany’s Excellence Strategy (EXC-2123 QuantumFrontiers Grants No. 390837967), through CRC 1227 (DQ-mat) within Projects No. A05 and through the QuantERA 2025 (QPR) co-funded project eBer-26-37748 (INTEGRA), and the German Space Agency (DLR) with funds provided by the German Federal Ministry for Economic Affairs and Climate Action (BMWK) (German Federal Ministry of Education and Research (BMBF)) due to an enactment of the German Bundestag under Grants No. DLR No. 50WM2450A (QUANTUS-VI), No. 50WM2545A (CAL-III), 50WM1552-1557 (QUANTUS-IV- Fallturm) and 50WM2677A (MARGE). JG and NG acknowledge funding from the EU project CARIOQA-PHB.
The authors acknowledge fruitful discussions with R.Walser on effect of anharmonicities regarding matter-wave lenses.

\bibliography{bibliography}
\bibliographystyle{apsrev4-2}

\end{document}